\documentclass[a4paper]{article}
\pdfoutput=1

\usepackage{graphicx}
\usepackage[utf8]{inputenc}
\usepackage{amsmath}
\usepackage{cases}
\usepackage{enumerate}
\usepackage{amssymb}
\usepackage{amsfonts}
\usepackage{mathrsfs}
\usepackage{amsthm}
\usepackage[font=footnotesize,margin=3mm]{caption}
\usepackage{subcaption}
\usepackage{textcomp}
\usepackage{bbm}
\usepackage[colorlinks=true, linkcolor=blue, citecolor=blue, urlcolor=blue]{hyperref}
\usepackage{booktabs}
\usepackage{authblk}
\usepackage{xcolor}

\newcommand{\figpath}{.}
\newcommand{\ud}{\mathrm{d}}

\newcommand{\Z}{\mathbb{Z}}

\newcommand{\pt}{\widetilde{p}}

\newcommand{\mut}{\widetilde{\mu}}

\newcommand{\Zt}{\widetilde{Z}}

\newcommand{\E}{\mathrm{E}}
\newcommand{\Var}{\mathrm{Var}}

\newcommand{\Bin}{\mathrm{Bin}}

\renewcommand{\Pr}{\mathbb{P}}

\newcommand{\One}{\mathbbm{1}}

\newcommand{\refEqn}[1]{\emph{Eq.~(\ref{eqn:#1})}}
\newcommand{\refFig}[1]{\emph{Figure~\ref{fig:#1}}}
\newcommand{\refTab}[1]{\emph{Table~\ref{tab:#1}}}
\newcommand{\refSec}[1]{\emph{Section~\ref{sec:#1}}}
\newcommand{\refApp}[1]{\emph{Appendix~\ref{sec:#1}}}
\newcommand{\refSubsec}[1]{\emph{Section~\ref{subsec:#1}}}

\newcommand{\given}{\;|\;}

\newcommand{\ml}{\overline{m}}

\newcommand{\mh}{\hat{m}}

\newcommand{\Ii}{I_{i}}
\newcommand{\Ji}{J_{i}}

\begin{document}

\title{Characterizing the Initial Phase of Epidemic Growth on some Empirical Networks}
\date{September 1, 2017}
\author[1,2]{Kristoffer Spricer}
\author[1]{Pieter Trapman}
\affil[1]{Department of Mathematics, Stockholm University, 106~91~Stockholm, Sweden}
\affil[2]{Corresponding author, email: \href{mailto:spricer@math.su.se}{spricer@math.su.se}}
\maketitle

\begin{abstract}
A key parameter in models for the spread of infectious diseases is the basic reproduction number $R_0$, which is the expected number of secondary cases a typical infected primary case infects during its infectious period in a large mostly susceptible population. In order for this quantity to be meaningful, the initial expected growth of the number of infectious individuals in the large-population limit should be exponential.

We investigate to what extent this assumption is valid by performing repeated simulations of epidemics on selected empirical networks, viewing each epidemic as a random process in discrete time. The initial phase of each epidemic is analyzed by fitting the number of infected people at each time step to a generalised growth model, allowing for estimating the shape of the growth. For reference, similar investigations are done on some elementary graphs such as integer lattices in different dimensions and configuration model graphs, for which the early epidemic behaviour is known.

We find that for the empirical networks tested in this paper, exponential growth characterizes the early stages of the epidemic, except when the network is restricted by a strong low-dimensional spacial constraint, such as is the case for the two-dimensional square lattice. However, on finite integer lattices of sufficiently high dimension, the early development of epidemics shows exponential growth.
\end{abstract}

\hspace{10pt}


\pagebreak

\section{Introduction}
\label{sec:introduction}
A key parameter in many mathematical models that describe the spread of infectious diseases is the basic reproduction number $R_{0}$. It may be understood as the expected number of other individuals a typical infected individual infects during his/her infectious period in a large mostly susceptible population \cite[page 4]{DiekmannEtal2013}. The basic reproduction number serves as a threshold parameter, in the sense that in most standard models, if $R_{0} \leq 1$ a large outbreak is impossible, while if $R_{0}>1$, a large outbreak occurs with positive probability. Furthermore, in those models, preventing a fraction $1-1/R_{0}$ of the infections (e.g.\ through vaccination) is enough to stop a major outbreak \cite[page 209]{DiekmannEtal2013}.

The above properties of $R_{0}$ are strongly connected to the correspondence of $R_{0}$ with the offspring mean of a branching process approximation of the epidemic. So for $R_{0}$ to be meaningful, the initial expected growth of the number of infectious individuals in the large-population limit should be exponential. This exponential growth is present in SIR epidemics (Susceptible $\to$ Infectious $\to$ Recovered; a definition is given in \refSubsec{epidemics}) in large homogeneously mixing populations, in which all individuals have the same characteristics and all pairs of individuals independently make contacts with the same rate. This exponential growth is also present in many well-studied generalizations of this SIR model in large homogeneously mixing populations. Generalizations are possible by leaving the SIR framework and allow for SIS (Susceptible $\to$ Infectious $\to$ Susceptible) or SIRS (Susceptible $\to$ Infectious $\to$ Recovered $\to$ Susceptible) models, or models with demographic turnover through births, deaths and migration. Other generalizations are allowing for heterogeneity among the individuals and contact rates between pairs, e.g.\ through allowing for household structures, multi-type structures and some network structures in the population (see e.g.\ \cite{DiekmannEtal2013} for descriptions of these  models and population structures). Even with these generalizations, major outbreaks of epidemics still show exponential growth in the initial phase of the epidemic and therefore $R_{0}$ is a meaningful parameter (see \cite{Trapman2016} and references therein).

A trade off between realism and analytical tractability is often necessary in developing a mathematical model. Because of the reasons stated above, in many instances this tractability requires the possibility of exponential growth in the model, either directly or as a byproduct of other assumptions. It is not a-priori clear in which cases real-life spread of infectious diseases allows for a meaningful definition of $R_{0}$ and in which cases the use of $R_{0}$ may be misleading and other key parameters should be estimated. For example, it is well known that SIR-epidemics on essentially 2-dimensional networks grow linearly whenever contacts between vertices are mostly local, i.e.\ if the probability of long range contacts decays sufficiently fast. The epidemic then spreads in the form of travelling waves on the plane (see e.g.\ \cite{Mollison1977, Wallace1991}, but also \cite{Grassberger2013, Trapman2010} for models where long-range contacts change the behaviour of the spread). Human physical activity is mostly restricted to the 2-dimensional nature of the earth's surface and a natural assumption is that graphs based on human interactions (e.g.\ in social networks) may also show this restriction.

In the present paper we study (simulated) epidemics on several theoretical and empirical networks and investigate to what extent they exhibit exponential growth and to what extent they exhibit subexponential growth. All, but one, of the empirical networks we use are taken from \cite{StanfordData} and we are aware that those networks are at best a proxy for networks relevant for the spread of infectious diseases. However, it is hard, if not impossible, to obtain complete network data for more relevant networks. Although we focus on the possibility of exponential growth of epidemics on networks, our interest goes beyond epidemics and epidemics on networks mainly serve as an example of of a stochastic process, for which the analysis strongly depends on implicit assumptions made on the network. We expect that our discussion on the stochastic behaviour of epidemics on empirical networks and on stochastically generated networks also apply to rumours, evolution or games on the networks. Furthermore, quantities of interest such as the diameter and typical distances in networks \cite[Chapter 1]{Hofs14a} are related to the possibility of exponential growth of an epidemic on the network, and in this context the typical distances are also strongly related to the so-called ``six degrees of separation'' and ``small-world'' phenomena \cite{watts1998collective}.

We study the spread of SIR epidemics on graphs/networks in discrete time (see \refSubsec{epidemics} below). In the graphs \emph{vertices} represent people and \emph{edges} represent relationships between people through which the infectious disease may spread. For any given vertex the vertices that can be reached directly through an edge are called the \emph{neighbours} of the vertex. The number of neighbours of a vertex is called the \emph{degree} of the vertex. We assume that the infection can spread in both directions on any edge, i.e.\ the graph is not directed.
Throughout, we assume that the epidemic starts with a single \emph{infected} vertex, the generation 0 vertex (called the \emph{index case}), while all other vertices are \emph{susceptible} to the disease. In each time step the infected vertices infect a subset of their susceptible neighbours, according to some probabilistic law (discussed below), after which they \emph{recover} and become immune forever. In the next time step the newly infected vertices (the generation 1 vertices) can infect their susceptible neighbours and so on. The epidemic ends when there are no more infected vertices. 

Several properties of the network can affect the basic reproduction number, both local (such as the degree of a vertex) and global structural properties.
\begin{figure}[ht]
	\centering
	\includegraphics[width=0.48\textwidth] {\figpath/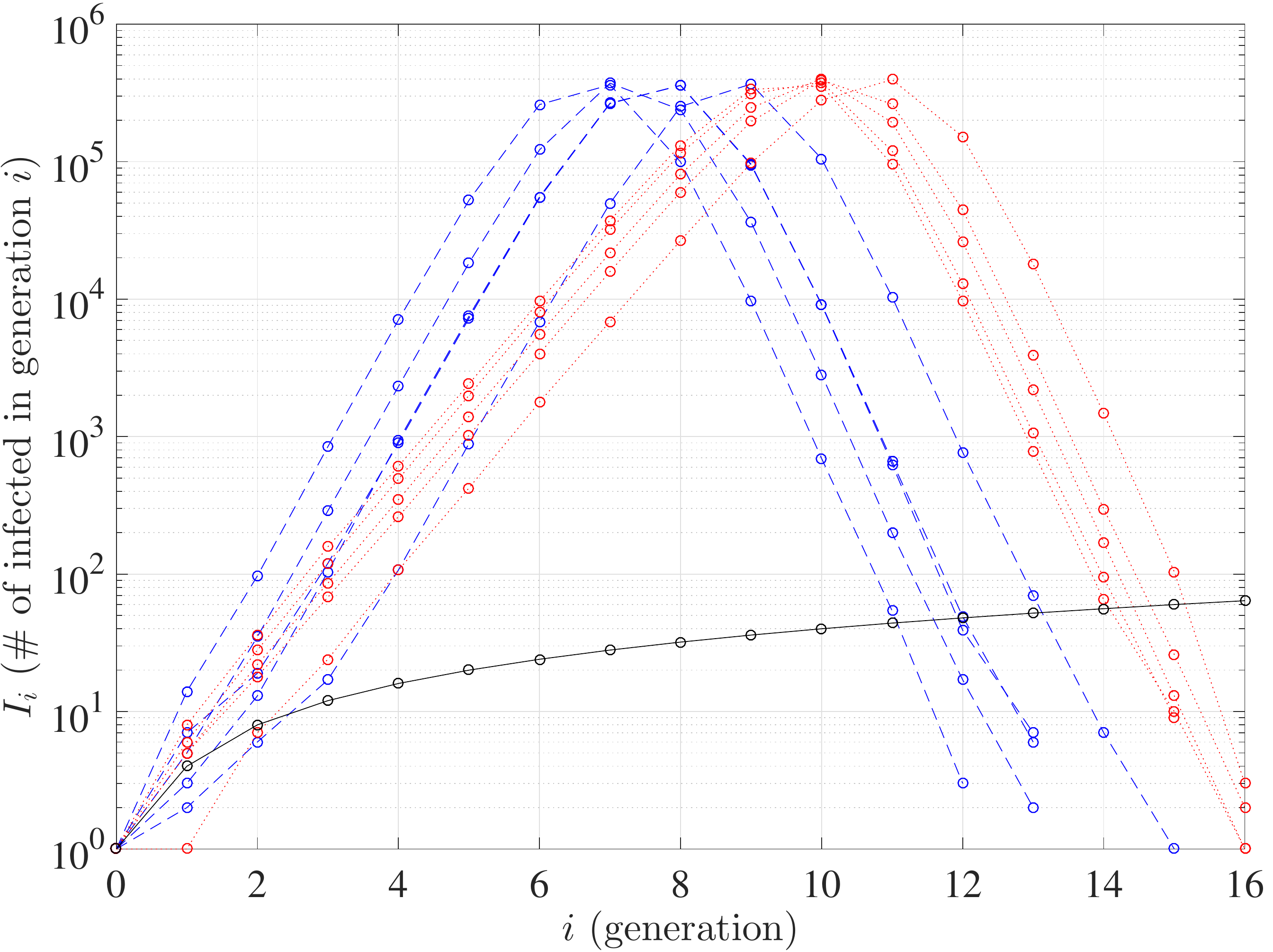}
	\includegraphics[width=0.48\textwidth]{\figpath/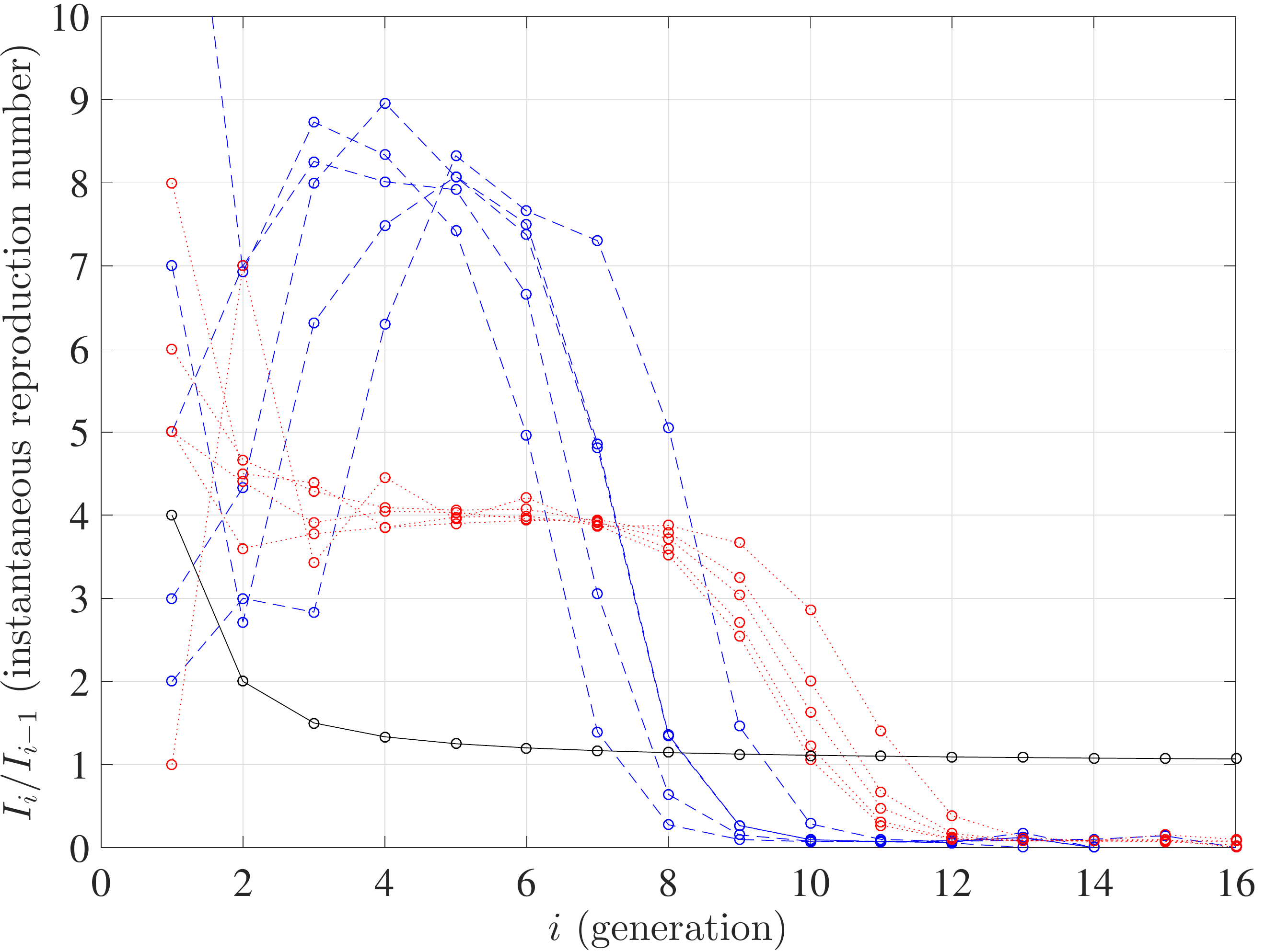}
	\caption{The number of infected vertices (left) and the instantaneous reproduction number (right) as a function of the epidemic generation for three network models---the square lattice (solid black line), the Erd{\H o}s-Rényi model (dotted red line) and the configuration model with a geometric degree distribution with weight on 0 (dashed blue line)---all with mean degree 4. For each plot the size of the population is $10^6$ vertices. In each generation infected vertices infect all of their susceptible neighbours. With the chosen axes scaling in the left plot, the data points fall on a straight line if the growth is exponential, corresponding to an approximately constant instantaneous reproduction number in the right plot.}
	\label{fig:ExampleEpidemics}
\end{figure}
As an example of the relevance of those structural properties, in \refFig{ExampleEpidemics} we have plotted a few SIR-epidemics that were simulated on three different network types that all have the same mean degree~4. These networks are described in more detail in \refSubsec{graphbasics}. Here we have assumed that infected vertices infect all of their susceptible neighbours. We observe, that although the three graphs have the same mean degree, epidemics on the three graphs develop differently. For the \emph{Erd{\H o}s-Rényi} model and the \emph{configuration model} early in the epidemic the growth is approximately exponential, while the \emph{square lattice} exhibits essentially linear growth. In the right figure for the first two graphs this is illustrated by approximately constant \emph{instantaneous reproduction numbers} (the average number of new infections caused per infected vertex in the current infection generation), well above 1 in the early parts of the epidemic, after which the instantaneous reproduction number drops to a value below 1. For the square lattice model the instantaneous reproduction number drops rapidly from the very start of the epidemic, asymptotically approaching 1. In the latter case the subexponential development is an effect of the spatial structure of the network.

Before starting the formal introduction and analysis of our models, we have to discuss some ideas behind stochastic models of real or simulated epidemics. Much of the theory on epidemics on graphs is about obtaining results for epidemics on infinite graphs, such as Euclidean lattices \cite{Grimmett99}, or obtaining asymptotic results for a sequence of related epidemics on finite graphs, when the graph size grows to infinity (e.g.\ \cite{BallEtal1995}). 
For example, for a configuration model graph (discussed more in \refSubsec{graphbasics}) with $n$ vertices, the initial growth of an epidemic can be analyzed using a branching process approximation \cite{Jagers75, BallEtal1995, BallSirlTrapman09}, which can be shown to be exact (under some extra conditions) until roughly $\sqrt{n}$ vertices have been infected, with probability tending to 1 if $n \to \infty$. This follows from a \emph{birthday problem} type argument (see e.g.\ \cite[page 54]{DiekmannEtal2013}).

For branching processes it is known that, if the branching process survives, the growth is almost surely asymptotically exponential (see \cite{Jagers75}), and the instantaneous reproduction number converges to its expectation, which is the basic reproduction number $R_{0}$. 
If a substantial fraction of the vertices have been infected in an SIR epidemic on a finite graph, with high probability, some neighbours of newly infected vertices have already been infected before, thus reducing the instantaneous reproduction number. Exponential growth, if it exists, is thus only visible in the early phases of the epidemic, when only a small portion of the graph has been infected.

In our analysis, we treat the empirical networks as if they are realizations of some unspecified random graph model, which can be defined for an arbitrary large number of nodes in the network. By studying epidemics on these realizations we try to answer whether, in the large population limit of the random graph, exponential growth is possible. However, we only have access to a limited number of such realizations and we cannot freely control the number of nodes in them. Also the real dynamics through which an empirical network is created are probably too hard to describe and analyze and possibly not even random. Therefore mathematical models and results based on such empirical realizations should be interpreted with care. Because there are infinitely many random graph models of which the empirical graphs may be a realization, we need to make more assumptions on the models behind the empirical graphs.

Our key assumption is that if the number of vertices in this random graph goes to infinity and if exponential growth is possible, then at some level (see \cite{BallSirlTrapman09}) a branching process approximation is possible and $R_{0}$ is well defined. This $R_{0}$ should then be estimated from the initial phase of a (simulated) epidemic on the finite empirical network. We analyze the development of the epidemic during the first generations (with exception of the first generation) to see if it is consistent with this assumption.

In most of the analyzed empirical networks the unrestricted epidemic, where an infected vertex infects \emph{all} susceptible neighbours, grows so fast that the early phase of the epidemic is over in as few as 3-4 generations. To be able to study the epidemic for more generations, we restrict it by two different methods described in \refSubsec{restrictgrowth}. These restrictions reduce the reproduction number so that it takes longer before a substantial portion of the population has been infected. Since such restrictions could affect the way in which the epidemic spreads in addition to just slowing it down, a secondary objective is to study such effects.

The analysis of the graphs is performed by fitting the early portion of simulated individual epidemics to a \emph{generalized growth model} (see \refSubsec{estgrowth}). From the fit the shape of the growth can be estimated and this gives information on how well the growth conforms to exponential or to subexponential growth. The process is repeated for many simulated epidemics on each graph and the collective information from the analysis of these epidemics is used to compare the initial growth of epidemics on different graphs (see \refSec{results}). A discussion can be found in \refSec{discussion}.

\section{Model}
\label{sec:model}
In this section we present the models for epidemic growth on graphs that we exploit in the paper, expanding on some of the concepts introduced in \refSec{introduction}. We start with definitions and results concerning graphs in \refSubsec{graphbasics}. An overview of the specific graphs that we analyze in this paper is given in \refSubsec{dataset}. In \refSubsec{epidemics} we give a brief account of the SIR-model on graphs in \emph{discrete time} (i.e.\ in a generation perspective) and discuss what we mean by the growth of the epidemic, specifically by exponential growth. We describe the two methods we use to ``slow the epidemic down'' in \refSubsec{restrictgrowth}. The method used to analyze the growth of the simulated epidemics is presented in \refSubsec{estgrowth}.

\subsection{Graphs}
\label{subsec:graphbasics}
A finite graph is a set of $n$ vertices together with a set of edges that join vertices pairwise. We consider \emph{simple} and \emph{undirected} graphs, i.e.\ there are no self-loops (an edge connecting a vertex to itself) and no parallel edges (several edges join the same pair of vertices) and all edges are undirected \cite{Hofs14a}. As already mentioned, the \emph{degree} of vertex $v$ is the number of neighbours of $v$. We denote this degree by $d_v$.
For a given graph we talk about the degree sequence $\mathbf{d}=(d_{1}, d_{2}, \dots, d_{n})$. Without loss of generality we restrict the analysis to graphs where $d_{i}\geq 1$ for all $i=1, 2,\dots ,n$, since vertices with degree 0 cannot interact in an epidemic and are therefore not really part of the network. Let $Z$ be the degree of a vertex selected uniformly at random from the graph and define $p_{k}=\Pr(Z{=}k)$. The distribution of $Z$ is the \emph{degree distribution} of the graph and we define $\mu = \E[Z]$ and $\sigma^{2}=Var(Z)$.

Some graphs are \emph{deterministic}, e.g.\ the graph on the Euclidean integer lattice $\Z^{\eta}$, where $\eta$ is the dimension of the lattice. 
Here the integer points in $\Z^{\eta}$ are the vertices and edges exist between all pairs of vertices with Euclidean distance 1. We consider finite subsets (tori) of the infinite graph $\Z^{\eta}$ in order to make comparisons with other finite graphs.
Other graphs are \emph{random} in the sense that they are constructed probabilistically, e.g.\ the Erd\H{o}s-Rényi graph (\cite{ErdosEtal1959,Durrett2007}) and configuration model graphs (\cite{MolloyEtal1995,Durrett2007}). In both cases the number of vertices is given and finite.

In the Erd\H{o}s-Rényi graph there exists an edge between any two vertices with probability $\frac{\lambda}{n-1}$, where $\lambda$ is a given constant equal to the expected degree of a vertex selected uniformly at random, and the presence or absence of possible edges are independent. This construction results in the degree distribution of a vertex selected uniformly at random being the binomial distribution with parameters $n-1$ and $\frac{\lambda}{n-1}$. For $n\to\infty$, the degree distribution converges to a Poisson distribution with parameter $\lambda$.

In the configuration model graph, we either start with a given degree sequence $\mathbf{d}$ (which may be taken from an empirical network) for the vertices, or the degree sequence is independent and identically distributed (i.i.d.), with given distribution $D$ (which may also be taken from an empirical network). We then create the graph as follows: To each (for the moment unconnected) vertex we assign a number of ``stubs'' corresponding to its degree.
The stubs are paired uniformly at random to create edges. Any left over stub is deleted and so are any self-loops, while parallel edges are merged to one edge. So, the created graph is simple. If the degree sequence is an i.i.d.\ sequence with distribution $D$, and if $D$ has finite mean, then in the limit $n\to\infty$ the degree distribution of the obtained configuration model graph converges in probability to the distribution $D$ \cite{britton2006generating}.

\emph{Empirical graphs} are created from real world data, e.g.\ from observed social interactions within a group of people. 
These networks have a given finite size, while we are interested in asymptotic results when the population size $n \to \infty$. 
Therefore, as pointed out in \refSec{introduction}, we analyze (processes on) the empirical graph as if the empirical graph is a realization of a random graph, which can be analyzed for $n \to \infty$. The exact mechanism of constructing the random graph is typically unknown. The empirical graphs we have analyzed in this paper are described in \refSubsec{dataset}, below.

\subsection{The studied networks}
\label{subsec:dataset}
In this subsection we present the networks we use to generate the graphs that are analyzed in this paper and we discuss some of their properties.

A summary of properties of the networks used in this paper can be found in \refTab{investigated}. The first three networks in the table are from the Stanford Large Dataset Collection (\cite{StanfordData}).
\begin{table}[ht]
	\centering
	\begin{tabular}{@{}lrrl@{}}
		\toprule
		Data set & \# vertices & \# edges & Type of graph\\
		\midrule
		soc-LiveJournal1	& 3\,823\,816 & 25\,624\,154 & online social network\\
		ca-CondMat			& 23\,133	& 93\,439		& scientific collaboration network\\
		roadNet-PA			& 1\,088\,092 & 1\,541\,898	& road network in Pennsylvania\\
		Swedish population & 7\,616\,569 & 18\,139\,894 & workplace and family\\
		D2							& 1\,000\,000 & 2\,000\,000 & 2-dimensional lattice\\
		D6							& 1\,000\,000 & 6\,000\,000 & 6-dimensional lattice
	\end{tabular}
	\caption{An overview of networks that are investigated in this paper. The number of undirected edges and the number of vertices with at least one edge are indicated.}
	\label{tab:investigated}
\end{table}

The graphs are discussed below.

\begin{itemize}
	\item \textbf{soc-LiveJournal1} is a large online social network that allows for the formation of communities. On the network people state who their ``friends'' are and although this does not have to be mutual, it often is. In our model we only consider the mutual statements of friendship and let these be represented by undirected edges, while people are represented by vertices. In figures the graph is referred to as ``LJ''.
	\item \textbf{ca-CondMat} is based on the \emph{arXiv} condensed matter collaboration network (COND-MAT). Authors are represented as vertices and undirected edges are present between all authors that are listed as co-authors of the same paper. In figures the graph is referred to as ``CM''. 
	\item \textbf{roadNet-PA} is based on the road network of Pennsylvania. Intersections between roads are represented by vertices and roads are represented by edges. Because of the spatial nature of a road network we expect to see spatial restrictions in this network and this is why it was included in the analysis. In figures the graph is referred to as ``Rd''.
	\item \textbf{Swedish population}\footnote{Data kindly supplied by Fredrik Liljeros, Department of Sociology, Stockholm University} is a large network that is based on data containing only the workplace and family affiliation of people in Sweden (see also \cite{Holm2002sverige}). Although the used dataset does not contain geographic information, it may still be assumed that family location and workplace location can be spatially correlated, thus imposing a spatial structure on the entire graph.

	Because some of the workplaces are large, we have assumed that people interact with colleagues only in smaller working groups. We model this by (randomly) dividing the workplaces into groups of 7 people (with at most one group in each company having a size between 1 and 6 when the company size is not divisible by 7).
	
	A \emph{reference} version of this dataset was also tested where company affiliation was assigned at random to each vertex, while keeping the distribution of workplace sizes fixed. If epidemics on this reference graph differ from the original graph, this could be an indication of spatial restrictions on the original graph. In figures the graphs are referred to as ``Sw'' and ``SR'', respectively.
	\item \textbf{D2} is a finite regular square lattice (on $\Z^{2}$) in the shape of a torus with sides of $10^{3}$ vertices, thus in total $10^{6}$ vertices. Because of the torus shape there is no center in the graph and the development of the epidemic does thus not depend on where the epidemic starts. In figures the graph is referred to as ``D2''.
	\item \textbf{D6} is a finite regular lattice on $\Z^{6}$ in the shape of a torus with sides of 10 vertices, thus in total $10^{6}$ vertices. As for the D2-graph there is no center in the graph. In figures the graph is referred to as ``D6''.
\end{itemize}

\subsection{Epidemics on Graphs}
\label{subsec:epidemics}
In the context of epidemics on graphs we think of the vertices as people and of the edges as relationships by which infected people can infect other people. Because the graphs are undirected, the epidemic can spread in both directions along any edge. We consider SIR-epidemics, where each vertex is either \emph{susceptible}, \emph{infected} (and infectious) or \emph{recovered}. A vertex that is recovered is immune and can never be infected again. In this paper we restrict the analysis to epidemics in \emph{discrete time} and also assume that each infected vertex stays infected for only one time unit before it recovers and cannot spread the infection further. This model corresponds to the so-called Reed-Frost model on graphs \cite[p.48]{DiekmannEtal2013}. The above implies that, for any finite graph, the epidemic must eventually end when there are no more infected vertices left. The total number of vertices that have been infected during the course of the epidemic is called \emph{the final size} of the epidemic.

The first vertex to be infected is called the \emph{index case}. We assume that the index case was infected at time $i=0$, where time represents the generation number. The index case then spreads the infection to (a subset of) its neighbours and they in turn spread it to (a subset of) their neighbours. For each generation $i$, we keep track of $\Ii$, the number of infected vertices in generation $i$, and of $\Ji=\sum_{k=0}^{i}I_{k}$, the total number of infected vertices up to and including generation $i$. Note that, \emph{if} the infection always spreads to \emph{all} neighbours, $\Ji$ is equal to the number of vertices within graph distance $i$ from the index case.

A measure of the rate at which the epidemic is growing is the \emph{instantaneous reproduction number} at time (generation) $i$, which we define as the \emph{average} number of offspring of a vertex in generation $i{-}1$
\begin{equation}
	\overline{m}_{i}=\frac{\Ii}{I_{i-1}},
\end{equation}
for $i{\geq}1$ and conditioned on $I_{i-1}{>}0$. The instantaneous reproduction number depends on how many neighbours an infected vertex has, on how many of the neighbours that are still susceptible and on the mechanism by which the vertex infects its neighbours. In the early phase of an epidemic $\overline{m}_{i}$ may be approximately constant as a function of $i$, but for a finite graph it must eventually decrease as there are fewer and fewer susceptible vertices left. If vertices have different degrees or if vertices have different local environments, then the development of the epidemic also depends on which vertex is the index case. 

On the square lattice, $\Z^{2}$, the growth of an unrestricted epidemic (i.e.\ when an infected vertex infects \emph{all} susceptible friends) is initially linear with $\Ii=4i$ and $\ml_{i}=\frac{i}{i-1}$, $i\geq 1$ (conditioned on $I_{0}=1$).

On a configuration model graph the early part of the epidemic (if the first generation is ignored) is well approximated by a \emph{Galton-Watson} branching process in discrete time, where all individuals reproduce independently with offspring distribution $X$, with $m=\E[X]$. For a Galton-Watson process
\begin{equation}
	\Ii=\sum\limits_{j=1}^{I_{i-1}}X_{i,j},
\end{equation}
where $X_{i,j}$ are all independent and distributed as $X$. We see that
\begin{equation}
\E\left[\Ii\given I_{i-1}\right]=I_{i-1}\E[X]
\end{equation}
and that
\begin{equation}
\Var\left[\Ii\given I_{i-1}\right]=I_{i-1}\Var[X]
\end{equation}
so that both the (conditional) expectation and the (conditional) variance of $\Ii$ are proportional to the size of the previous generation. We use these relationships in the analysis of the data, see \refSubsec{estgrowth}. 

In the large population limit the expected reproduction number
\begin{equation}
 \E[\ml_{i}\given I_{i-1}{>}0]=m
\end{equation}
does not depend on the epidemic generation. The expected size of the $i$-th generation of the epidemic is $\E[\Ii]=m^{i}$ (given a single index case and assuming the branching process is valid from the first generation) and we see that the growth of the epidemic is exponential if $m>1$.

As shown in \cite[page 36]{Guttorp1991}
\begin{equation}
	\label{eqn:mhat}
	\mh_{i}=\frac{\Ji-1}{J_{i-1}},
\end{equation}
(assuming that $I_{0}=1$) is a better estimator for $m$ than $\ml_{i}$. If the branching process approximation is valid first from generation 2 then we can modify \refEqn{mhat} slightly to obtain
\begin{equation}
	\label{eqn:mhatp}
	\mh_{i}=\frac{\Ji-J_{1}}{J_{i-1}-J_{0}}.
\end{equation}
This latter expression is most relevant in this paper, since we select the index case uniformly at random among all vertices, while subsequent vertices are infected by following edges from an infected vertex. This causes the degree distribution of the index case to differ from that of vertices that are infected later, as we explain now (see also \cite{Newm02}).

Before continuing, we remind the reader that $\mu=E[Z]$ and $\sigma^{2}=\Var(Z)$, where $Z$ is the degree of a vertex that is chosen uniformly at random among all vertices in the graph (see also \refSubsec{graphbasics}). Let $\Zt$ be the degree of a vertex that is selected by first selecting an edge uniformly at random and then selecting one of the two connected vertices at random. On the configuration model graph, again ignoring the first generation, initially and for as long as the branching process approximation is valid, the epidemic growth is governed by $X\sim\Zt{-}1$, where
\begin{equation}
	 \pt_{k}=\Pr\left(\Zt{=}k\right)=\frac{kp_{k}}{\mu}.
\end{equation}
The ``$-1$'' is because the infection cannot spread back to ``the infector'' since it is by definition not susceptible any more. The \emph{expected} reproduction number in the early stages of the epidemic is thus $m=\mut{-}1$, where
\begin{equation}
	 \mut=\E\left[\Zt\right]=\frac{\E\left[Z^{2}\right]}{\E[Z]}=\mu+\frac{\sigma^{2}}{\mu}.
\end{equation}
Thus $\mut$ can be much larger than $\mu$ if $\sigma^{2}$ is much larger than $\mu$.

\subsection{Restricting the Reproduction Number}
\label{subsec:restrictgrowth}
Similar to the configuration model, some of the empirical graphs analyzed in this paper early on in the epidemic have instantaneous reproduction numbers that are much larger than the mean degree of the graph. An unrestricted epidemic on such a graph grows very fast, infecting most of the population in just a few generations. This makes it difficult to assess if the growth is exponential or not. To work around this problem, we restrict the epidemic so that it develops slower, giving more generations to analyze. This is reasonable, since in real world epidemics we do not expect that each infected vertex infects all of its neighbours.

We use two methods to restrict the instantaneous reproduction number:
\begin{enumerate}
	\item \emph{Maximum bound, with replacement}: Every infected vertex distributes $c$ infection attempts uniformly at random \emph{with} replacement among all neighbours (including the one who infected him). Here $c$ is a constant. Thus an infected vertex can infect $0$ (if all attempts are with non-susceptible neighbours) up to $c$ of its neighbours (if all attempts are with susceptible ones). If $c$ is sufficiently large (often $c=2$ is enough), this method (typically) allows for large epidemics to develop since both infected vertices with few and infected vertices with many neighbours have a good chance of infecting other vertices. The method is similar to, but not identical with the method used in \cite{MalmrosEtal2016}.
 
 Note that this method creates an asymmetry between vertices, in effect turning the undirected graph into a directed graph: it may be that if vertices $v_1$ and $v_2$ are neighbours then it is more likely that $v_1$ infects $v_2$ (should $v_1$ become infected before $v_2$), than that $v_2$ infects $v_1$ (should $v_2$ become infected first).
	\item \emph{Bernoulli thinning}: Each susceptible neighbour is infected with probability $p$. This method is equivalent to the discrete-time version of the Reed-Frost model, where it is assumed that each infected vertex infects each neighbour with probability $p$ (\cite[page 48]{DiekmannEtal2013}). This is closely related to bond percolation on the graph \cite{Grimmett99}.
 
A disadvantage of Bernoulli thinning is that, for the datasets that we analyze, in order to significantly slow down the epidemic $p$ has to be so low that vertices with few neighbours have a high probability of not infecting any other vertex. The epidemic is spread mainly through high degree vertices, resulting in fewer infected vertices and a smaller final size of the epidemic. Eventually this has a negative effect on the number of generations that can be used to estimate the instantaneous reproduction number, counteracting the intention of the Bernoulli thinning.
\end{enumerate}

When one of the above mentioned restrictions is applied, an infected vertex typically infects only a subset of its neighbours and the epidemic develops differently on each realization, even if it starts with the same index case. This introduces randomness even for epidemics on non-random graphs.

In this paper we have chosen to slow the epidemic down in such a way that we obtain a sufficient number of generations to analyze, while still leaving the possibility of having a large epidemic. For this purpose the maximum bound restriction $c=3$ worked on all graphs. We used this throughout the analysis, unless explicitly stated otherwise. On each graph, for the Bernoulli thinning we then select a value $p$ such that the average reproduction number, over many simulations, early on in the epidemic is close to that of epidemics restricted using maximum bound.

We also restrict the epidemics on the reference graphs. For the configuration model graphs we obtain exact expected reproduction numbers that are valid for as long as the branching process approximation holds. For calculating these expected reproduction numbers using the two restriction methods, we start with the offspring distribution $X$ and derive the expectation of the \emph{restricted} distribution $L$.

We remind the reader that $\Zt$ is the degree distribution of vertices reached early on in an epidemic on a configuration model graph, excluding the index case itself (see \refSubsec{epidemics}). For Bernoulli thinning, remembering that $X\sim\Zt{-}1$ in the configuration model, we then have that
\begin{equation*}
	L\given\Zt=k\sim \Bin(k-1,p)
\end{equation*}
so that
\begin{equation*}
	 \E\left[L\given\Zt=k\right]=(k-1)p
\end{equation*}
and
\begin{equation*}
	 \E\left[L\given\Zt\right]=\left(\Zt-1\right)p.
\end{equation*}
Thus
\begin{equation}
	\E\left[L\right] = \E\left[\E\left[L\given\Zt\right]\right]=\E\left[\left(\Zt-1\right)p\right]=\left(\mut-1\right)p=\left(\mu+\frac{\sigma^{2}}{\mu}-1\right)p.
\end{equation}
For the maximum bound restriction, in \refApp{derivationrepno} we derive that
\begin{equation}
	\E[L] = \E\left[\E\left[L\given\Zt\right]\right]=\E\left[(\Zt-1)\left(1-\left(1-\frac{1}{\Zt}\right)^{c}\right)\right].
\end{equation}
This expression can be simplified for the specific values of $c$ that we focus on in this paper:
\begin{equation}
	\E[L] = 
	\begin{cases}
		2-3\E\left[\frac{1}{\Zt}\right]+\E\left[\frac{1}{\Zt^{2}}\right] &\text{ when }c=2,\\[.4ex]
		3-6\E\left[\frac{1}{\Zt}\right]+4\E\left[\frac{1}{\Zt^{2}}\right]-\E\left[\frac{1}{\Zt^{3}}\right]&\text{ when }c=3.\\
	\end{cases}
\end{equation}
Using
\begin{equation*}
	\E\left[\frac{1}{{\Zt}^{n}}\right] = \frac{1}{\mu}\E\left[\frac{1}{Z^{n-1}}\right]
\end{equation*}
this can also be expressed as
\begin{equation}
	\E[L] = 
	\begin{cases}
		2-\frac{3}{\mu}+\frac{1}{\mu}\E\left[\frac{1}{Z}\right] &\text{ when }c=2,\\[.4ex]
		3-\frac{6}{\mu}+\frac{4}{\mu}\E\left[\frac{1}{Z}\right]-\frac{1}{\mu}\E\left[\frac{1}{Z^{2}}\right]&\text{ when }c=3.\\
	\end{cases}
\end{equation}

The branching process approximation of an epidemic on a configuration model graph together with the expected reproduction number of the restricted epidemic can be used as a reference for the empirical graphs.

\subsection{Estimating the Shape of Growth}
\label{subsec:estgrowth}
A direct method of analyzing the epidemic growth is to look at how $\Ii$ develops over the epidemic generations $i$. From the discussion of the expected growth shape for some graph types in \refSubsec{epidemics} we conclude that we need a method that is able to handle shapes from linear to exponential. One way to do this is to use a function that models the highest polynomial degree of the growth curve
\begin{equation}
	\Ii = \alpha(i+\beta)^{\gamma},
\label{eqn:largestpoly}
\end{equation}
where $\alpha$, $\beta$ and $\gamma$ are parameters that we determine by fitting the function to our data. In this paper $\gamma$ is the parameter of interest. We expect it to be close to 1 if the growth is linear (as for the square lattice) and it should be substantially higher than 1 if the growth is exponential (as for the configuration model). The parameter $\beta$ is introduced since we do not expect the first generation to show the same expected growth as subsequent generations (as discussed in \refSubsec{epidemics}). We thus ignore the first generation and allow for some offset for the time $i$. Unfortunately, because the parameters are highly dependent, the chosen parametrization in \refEqn{largestpoly} does not give good convergence when using standard methods of fitting the equation to data.

An alternative parametrization was originally suggested in \cite{Tolle2003} in the context of \emph{superexponential} growth and was used to study the impact of superexponential population growth on genetic variations in \cite{Reppell2014}. The same method was used for \emph{subexponential} growth in \cite{ViboudEtal2016}. The parametrization was developed for continuous time applications, but can be adapted to our discrete time data.

The basic idea in \cite{Tolle2003} is to start with a differential equation with two parameters
\begin{equation}
	\label{eqn:diffeq}
	\frac{\ud f(t)}{\ud t} = r f(t)^{a},
\end{equation}
where $f(t)$ can be viewed as modelling the total population size or the number of infected (depending on application) at time $t$ and $r$ and $a$ are parameters. $f(t)$ in continuous time corresponds to $\Ii$ in discrete time simply by setting $\Ii=f(i)$. $a$ defines the \emph{shape} of the growth curve, while $r$ is a proportionality constant which we may interpret as a measure of the \emph{rate} of growth. The solution to \refEqn{diffeq} depends on the value of $a$:
\begin{subnumcases}{f(t) =} 
	be^{rt} & \text{if }a=1 \label{eqn:f_exp}\\[.4ex] 
	\left(r(1-a)t+b^{1-a}\right)^{\frac{1}{1-a}} & \text{otherwise},\label{eqn:f_poly}
\end{subnumcases}
where $b = f(0)$ is given by the starting condition, the size of the population at time 0. We note that \refEqn{f_exp} is the limit of \refEqn{f_poly} as $a\to 1$. When $a=1$ we recognize that $r$ is the \emph{Malthusian parameter} (see e.g.\ \cite[page 10]{DiekmannEtal2013}). We also note that \refEqn{f_poly} is essentially a reparametrization of \refEqn{largestpoly}.

Taking $\Ii=f(i)$ in \refEqn{f_poly} we obtain
\begin{equation}
	\Ii = \left(r(1-a)i+b^{1-a}\right)^{\frac{1}{1-a}}. \label{eqn:I_poly}
\end{equation}

When performing the fit we take into account that the variance of $\Ii$ is not constant. Rather, we can expect it to increase if the generation size is larger and in our model we go further and assume that it is approximately \emph{proportional} to the generation size. This is reasonable considering that the conditional variance and the conditional expectation of the generation size are proportional to each other (as noted in \refSubsec{epidemics}). This lends itself well to a $log$-transformation to obtain
\begin{equation}
	\log(\Ii) = \frac{1}{1-a}\log\left(r(1-a)i+b^{1-a}\right).
	\label{eqn:I_log}
\end{equation}

Although this model has a singular point when the growth is exactly exponential ($a=0$), this case is unlikely with empirical data and we choose to ignore the singular point and use \refEqn{I_poly} as it is.

In the model, $\Ii$ is the data that is obtained from each individual simulated epidemic and $a$, $r$ and $b$ are treated as unknown parameters. By fitting \refEqn{I_log} to the data we obtain estimates of the parameter triple $(a, r, b)$. The fit is performed using least squares regression by supplying \refEqn{I_log} as a custom function to the \emph{fit}-function in \emph{Matlab} \cite{MatlabCFT}. The \emph{fit}-function is supplied with starting points $(0.5, 0.5, 0.5)$, minimum allowed values $(-10, 0, 0)$ and maximum allowed values $(5, 10^{5}, 100)$ for the parameter triple. In addition, $R_{adj}^{2}$ (the adjusted coefficient of determination, see e.g.\ \cite[page 433]{Tamhane2000}), produced by the \emph{fit}-function, was inspected, but the value was not used to discard any results. $R_{adj}^{2}$ were in general high, except for epidemics on the road network. These depart most from the shape assumed by the generalized growth model and this is also reflected in the large variation in parameter estimates that can be seen in \refFig{empirical_GenGrowth_Plot}.

Conditioned on having a \emph{good fit}, we can then interpret $a$ as a measure of how linear or how exponential the epidemic growth is. Values of $a$ close to 1 can be interpreted as having exponential growth, while values close to $0$ correspond to linear growth, such as we expect for the square lattice. Negative values correspond to sub-linear growth.

How good the fit needs to be to draw conclusions about a single epidemic depends on the application. However, through simulation we have access to many epidemics from each graph. Thus we can assess how similar or how different graphs are by comparing the parameter estimates from a large number of simulated epidemics for each graph. As already stated in \cite{ViboudEtal2016} this is a phenomenological approach and as such it does not properly justify why this specific model and parametrization of the growth curve should be used. We justify the method by also simulating epidemics on known (reference) graphs and by using parameter estimates from those. Our reference graphs are regular lattices and configuration model graphs. We interpret the parameter estimates from the empirical graphs with respect to those  obtained on the reference graphs.

The branching process approximation discussed in \refSubsec{epidemics} works well until there is a substantial probability that an infected vertex tries to infect an already immune vertex. Given that $J$ is the total number of vertices that has been infected in the epidemic, this probability would in a configuration model be approximately $\frac{J\mut}{n\mu}$, i.e.\ the proportion of already infected stubs divided by the total number of stubs. For the datasets we analyze this probability grows fastest for the LiveJournal dataset. This is because of the high quotient of $\frac{\mut}{\mu}\approx 5$. If we, arbitrarily, allow this probability to be at most 5\%, thus reducing the instantaneous reproduction number by approximately the same amount, we cannot allow $\frac{J}{n}$ to be more than approximately 1\% for the LiveJournal dataset and slightly higher for the other datasets. Setting this limit too low gives too few generations for the statistical analysis, thus increasing the confidence intervals for the parameters, and setting it too high means that the branching process approximation is no longer good and we should expect biased parameter estimates (generally too low values of $a$), indicating that the growth is not exponential), even when working with configuration model graphs.  In this paper we set the limit to 1\% for all datasets. We have tried (but not shown in this report) limits that are both lower and higher, and the chosen limit appeared to result in an acceptable compromise between imprecision and bias for the parameter estimates.

Finally, we make a couple of notes regarding the chosen model. First, we note that it is \emph{not} a predictive model, but rather a way to characterize the early phase of simulated epidemics on graphs. If we wanted to make predictions forward in time, then it is not certain that this model is the best method. We should then also validate the predictive properties of the model. Secondly, we are aware that data points are correlated, but we chose not to take this into account when fitting the data. We justify this by also including reference graphs in the same type of analysis that is used for the empirical graphs.

\section{Results}
\label{sec:results}
In this section we present results of the statistical analysis for epidemics on the empirical graphs and compare the result with epidemics on some reference networks.

We have used $10^{4}$ epidemics (with $I_{0}=1$) from each of the graphs in \refTab{investigated} and performed a least square fit to \refEqn{I_log}. Data used are from $i=1$ until approximately a total of $1\%$ of all vertices in the graph have been infected (see \refSubsec{estgrowth}). For the maximum bound restriction of epidemics, we set $c=3$ in the simulations. This is because some of the graphs do not allow for large epidemic outbreaks with $c=2$, while $c=3$ results in large epidemic on all graphs. For the corresponding simulation using Bernoulli thinning to restrict the epidemic, $p$ was selected to give a \emph{similar} growth rate early in the epidemic.

\begin{figure}[ht]
	\centering
	\includegraphics[width=0.49\textwidth]{\figpath/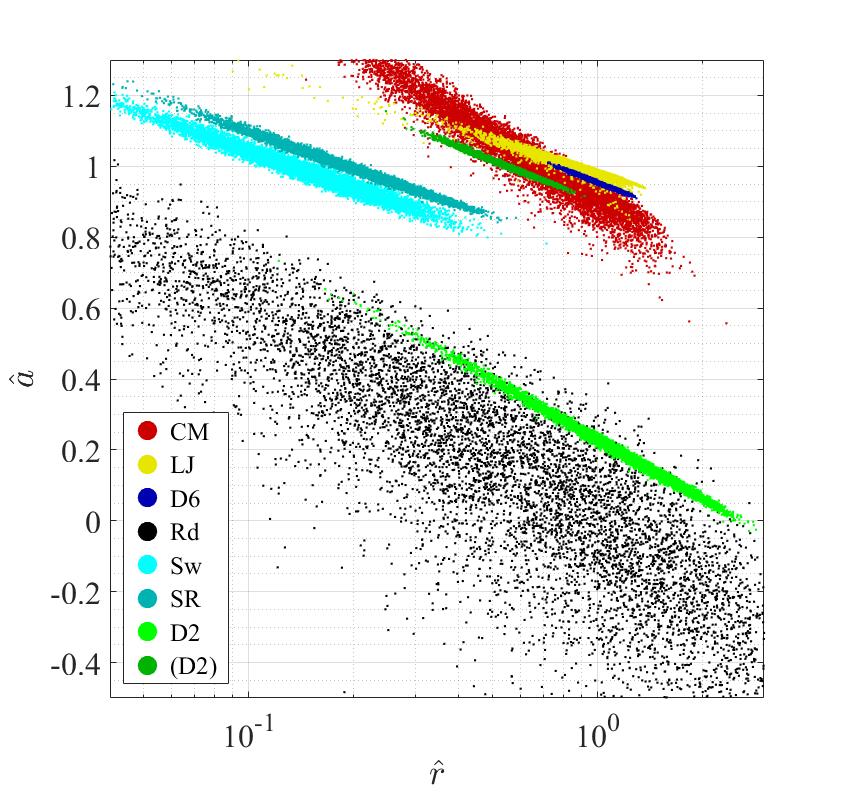}
	\includegraphics[width=0.49\textwidth]{\figpath/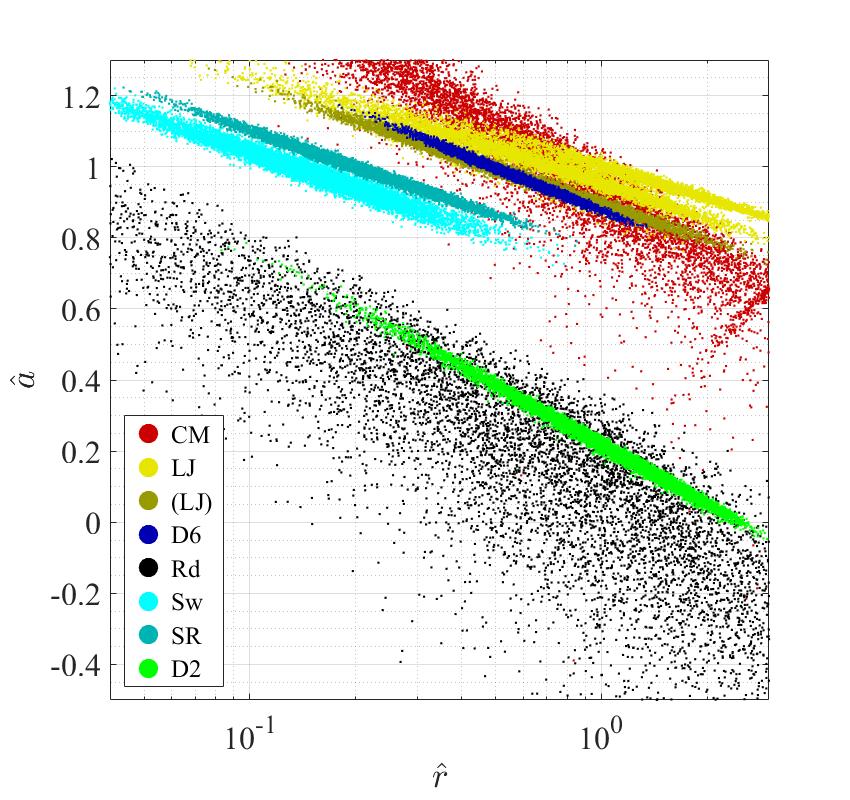}
	\caption{Each dot in the figure corresponds to an individual epidemic that has been fit to \refEqn{I_log}. For each graph $10^{4}$ epidemics were simulated. The overall ``cloud'' of point estimates of the parameters characterizes each graph in terms of what type of epidemics it produces. For the left figure maximum bound $c=3$ was used and for the right figure Bernoulli thinning was used. Graph names in parenthesis indicate that the configuration model was used. Note the logarithmic scale on the horizontal axis.}
	\label{fig:empirical_GenGrowth_Plot}
\end{figure}
The results are summarized in \refFig{empirical_GenGrowth_Plot} where we have plotted the estimated values of parameters $a$ versus $r$. We remind the reader that $a$ corresponds to the shape of growth where values close to 1 indicate exponential growth and values close to 0 indicate linear growth, while $r$ is a measure of the growth rate, corresponding to the \emph{Malthusian parameter} when we have exponential growth. In this paper the estimates of $a$ are of most interest. For reference, we have included some configuration model and square lattice graphs together with the empirical graphs.

We note that most of the graphs produce epidemics with estimated parameter values in the vicinity of $a=1$, while the road network and the square lattice data are spread out around $a=0$. This indicates that most of the graphs produce epidemics that grow exponentially early on, while the road network and the square lattice show an essentially linear growth. Note the similarity between the configuration model simulation of D2 (the square lattice) and some of the empirical graphs. Somewhat surprising to the authors is that D6 (the six dimensional lattice) seems to produce restricted epidemics that grow exponentially, while we would have expected polynomial growth for these (in this case with $a=4/5$, while the median estimated $a$-value is approximately 0.95).  The explanation is that because of the relatively high dimension of the graph and the strong restriction on the spread of the epidemic, early on vertices still have many available neighbours that are not yet infected and the epidemic can be approximated by a branching process. While this would eventually change to polynomial growth if allowed to continue long enough, there is no space for this in a finite graph. Note that low $r$ may also be a sign of non-exponential growth, since exponential growth with base close to 1 is hardly distinguishable from polynomial growth.

\begin{figure}[ht]
	\centering
	\includegraphics[width=0.49\textwidth]{\figpath/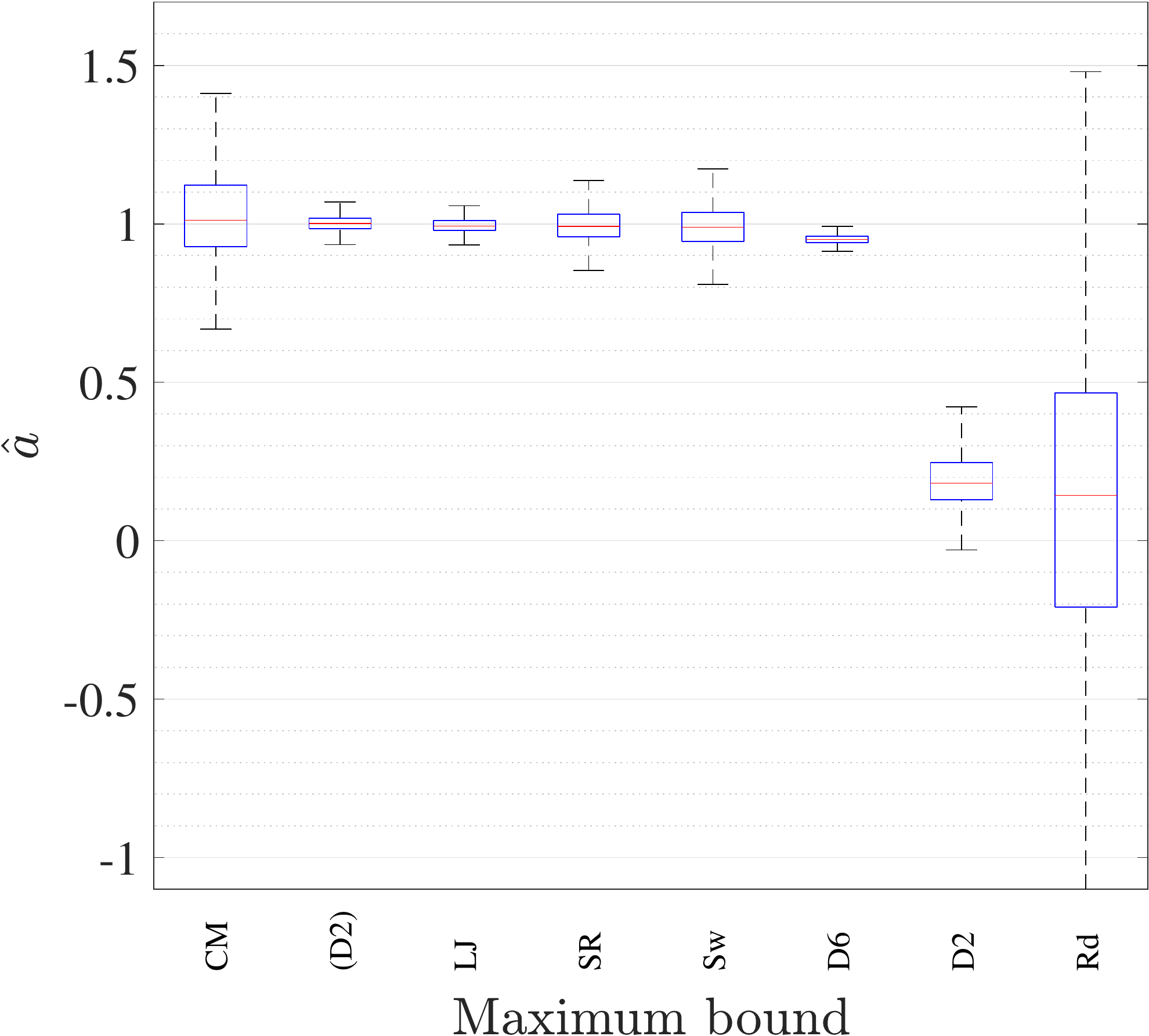}
	\includegraphics[width=0.49\textwidth]{\figpath/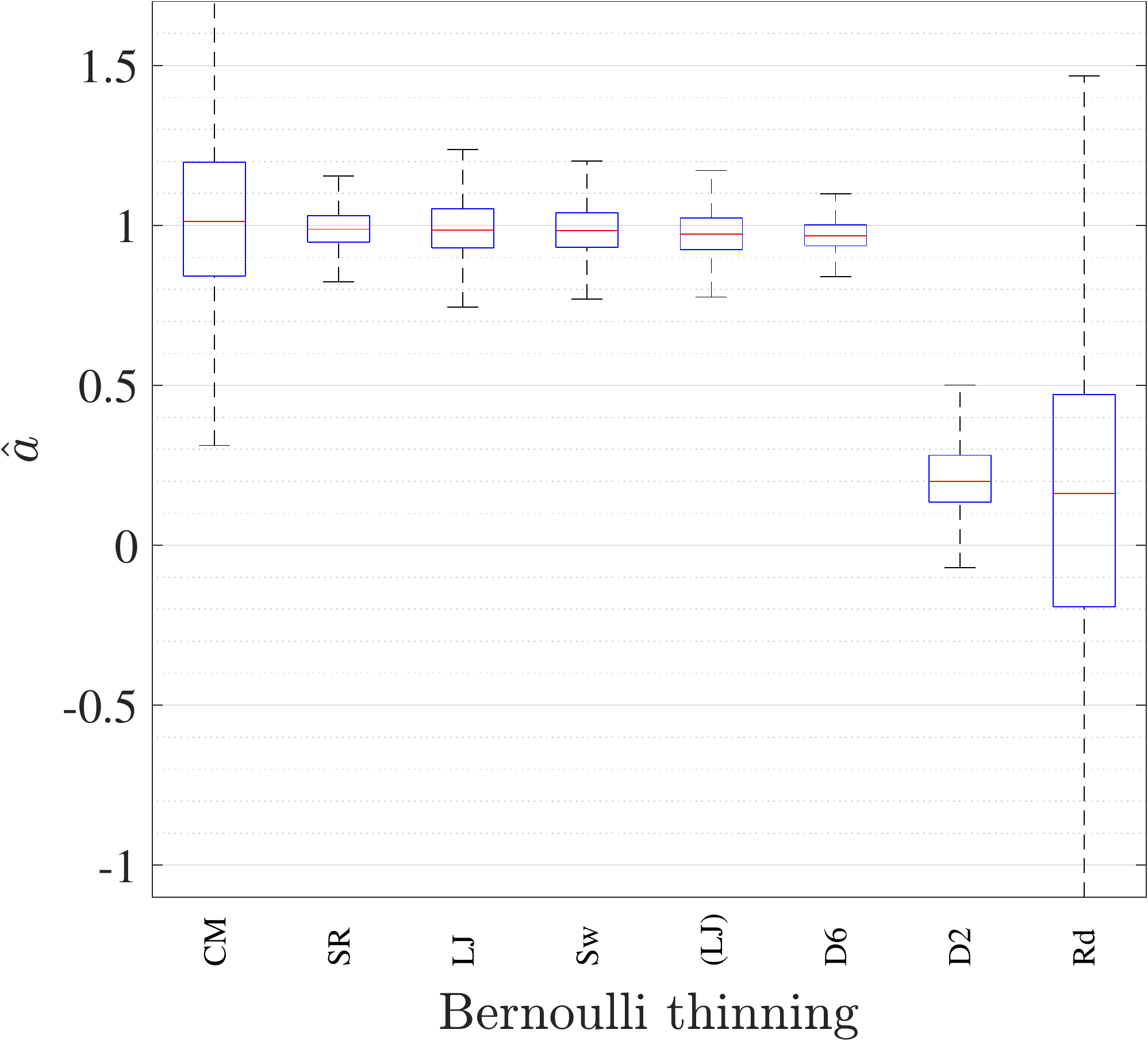}
	\caption{The figure shows a box plot of the estimates of $a$ for each graph from \refFig{empirical_GenGrowth_Plot}. Graph names in parenthesis indicate that the configuration model was used.}
	\label{fig:empirical_GenGrowth_Box}
\end{figure}
To better be able to observe differences in the estimated values of $a$, box plots of the $a$-estimates are shown in \refFig{empirical_GenGrowth_Box}. In the plot outliers have been ignored to make the central part of the data more visible.

\begin{figure}[ht]
	\centering
	\includegraphics[width=0.49\textwidth]{\figpath/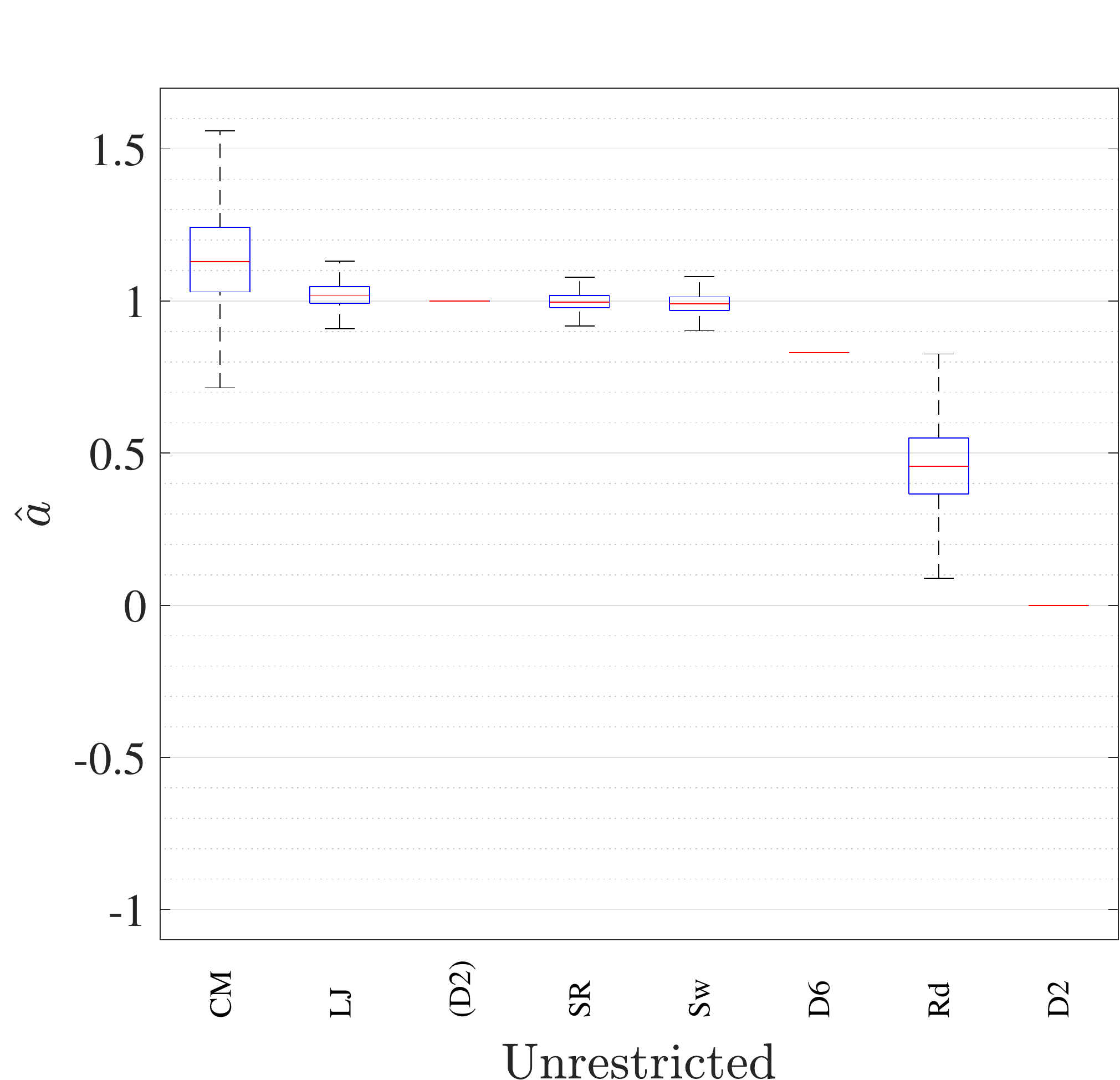}
	\includegraphics[width=0.49\textwidth]{\figpath/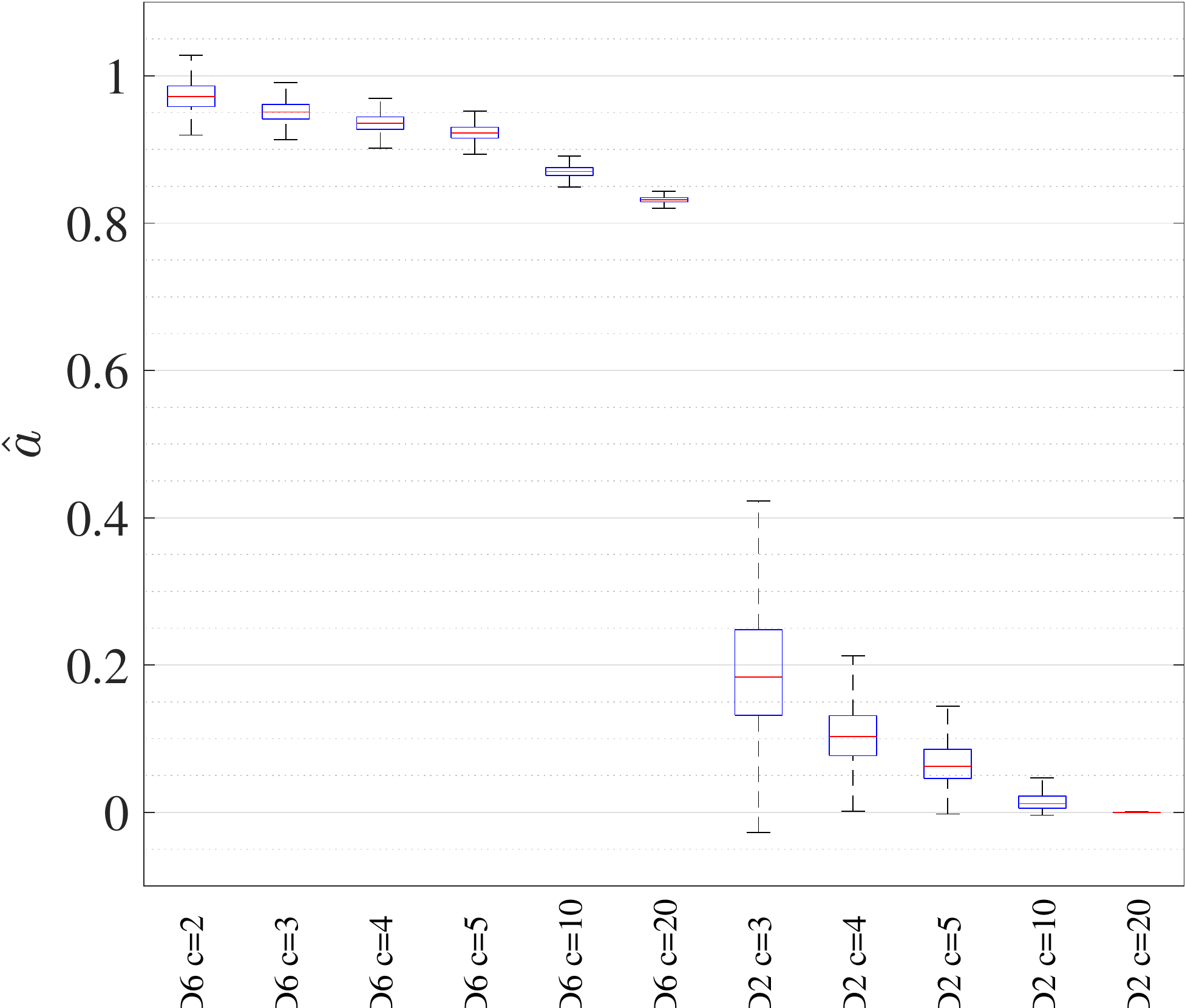}
	\caption{The figure to the left shows a box plot for unrestricted epidemics. Graph names in parenthesis indicate that the configuration model was used. The figure to the right shows epidemics restricted with different values of $c$.}
	\label{fig:empirical_GenGrowth_C}
\end{figure}
In order to see the effect of the restriction we place on the epidemic we show D2 and D6 using the maximum bound restriction, with different values of $c$, in \refFig{empirical_GenGrowth_C}. We note that lower values of $c$ (more restricted epidemic) move the parameter estimates towards higher $a$-estimate for both lattices. For the D2 graph event when we use the highest possible restriction $c=3$ the growth is still clearly polynomial, but for the D6 graph we can shift the $a$-estimates so close to 1 that epidemics on the graph appear exponential. We conclude that if the epidemic spreads through only some smaller fraction of the available edges we can see exponential growth early on in the epidemic. One underlying assumption for this conclusion is that large epidemics are possible in the first place, i.e.\ that the graph is sufficiently well connected.

From the plots we also see that epidemics on the road network show much more variation than on the square lattice. The road network seems to be a mixture of strongly connected portions and long stretches of vertices in long lines connected only by single edges along the way. This is what we may expect from a road network for a large geographical area consisting both of densely connected cities and loosely connected countryside.

\section{Discussion}
\label{sec:discussion}
The main purpose of this paper was to find a method to distinguish between empirical graphs which allow for initial exponential growth of an SIR epidemic and graphs which do not. If we know that exponential growth or close to exponential growth is possible, we can use statistical machinery already created for analyzing the growth potential of epidemics. To make this distinction we use the generalized growth model of \cite{ViboudEtal2016} as presented in \refSubsec{estgrowth} above. This model has three parameters, but  only $a$ (which describes the shape---polynomial or exponential---of the initial growth) and $r$ (which is a measure of the rate of growth) are relevant in this paper. We are mainly interested in $a$, but we cannot ignore $r$ because the estimates of the two parameters are strongly dependent. Indeed, we see in \refFig{empirical_GenGrowth_Plot} that although different epidemic simulations can produce very different parameter estimates, in the $(r,a)$ plot estimates of epidemics on different underlying networks can still be distinguished.

Ideally, when $a$ is close to 1 (how close depends on the application) we may 
conclude that the graph allows for epidemics that exhibit exponential growth. In (\refFig{empirical_GenGrowth_Box}) we visualize the distribution of the estimates for parameter $a$ for the individual graphs. This graph gives an indication of how close the growth of the epidemic is to exponential growth, but the figure must be interpreted with care. If the growth of the epidemic exactly follows the model with parameter $a =1$ and $r$ close to 0, then the growth is indeed exponential, but still very slow and it is very hard to distinguish this exponential growth from polynomial growth, with a larger $r$. Because the empirical networks are finite and we only observe a limited number of generations, 
we often do not have enough data to reliably distinguish between exponential growth with a small growth rate and polynomial growth. This observation is articulated in \refFig{empirical_GenGrowth_Plot}, where  we see that some simulated epidemics on the road network (which is clearly two dimensional) produce estimates of $a$ that are close to 1. However, for those simulations also the obtained estimates of $r$ are low (typically 0.1 or lower).

For the Swedish population dataset comparing the original dataset with a randomized version indicates that there are some effects that may be attributed to spatial constraints, but the difference is mainly seen on the rate of growth through the parameter $r$ and no so much on the parameter $a$. A possible conclusion is that the spatial constraints slow down the epidemic, but that the growth is still close to exponential (see e.g.\ \cite{Trapman2010} for a purely spatial model which allows for exponential growth of the epidemic).

The analysis of epidemics on the six dimensional lattice indicate that when the epidemic is restricted as in this report (\refSubsec{restrictgrowth}) the resulting early epidemic growth is essentially exponential. This can be explained as follows. Because vertices infect only a few of its neighbours, most neighbours of infected vertices will still be susceptible, so the local depletion of susceptibles is only felt after several  generations, when probably already a considerable fraction of all the vertices are no longer susceptible. In addition, on an infinite six dimensional lattice $I_n$ will grow as a five dimensional polynomial, which corresponds with an $a$-value of $4/5$ in \refEqn{I_poly}, which is relatively close to 1.

In the present work we only considered point estimates for the $(r,a)$ parameter pair, in future work it is worth studying confidence regions for those parameters, based on one single observed epidemic on a network. In addition to summarizing data by fitting it to a model, the strength of models is to be able to make predictions. There are two classes of predictions we might desire. We may want to predict the continued development of a single epidemic in the future based on how it developed up until some point in time. We may also want to predict the development of future (new) epidemics on the same graph based on knowledge of a (limited) number of previous epidemics. For these predictions it is essential that we know whether we may expect exponential growth or not. We have not attempted to investigate the possibility of making such predictions in this paper, but it is certainly worth studying in future work.

\section*{Acknowledgements}
P.T.\ was supported by Vetenskapsr{\aa}det (Swedish Research Council), project 201604566. The authors would like to thank Tom Britton for valuable discussions. 

\bibliographystyle{abbrv}
\bibliography{publicationsKSPT.bib}

\newcommand{\SortNoop}[1]{}
\begin{thebibliography}{10}

\bibitem{BallEtal1995}
F.~Ball and P.~Donnelly.
\newblock Strong approximations for epidemic models.
\newblock {\em Stochastic Process. Appl.}, 55(1):1--21, 1995.

\bibitem{BallSirlTrapman09}
F.~Ball, D.~Sirl, and P.~Trapman.
\newblock Threshold behaviour and final outcome of an epidemic on a random
  network with household structure.
\newblock {\em Adv. in Appl. Probab.}, 41(3):765--796, 2009.

\bibitem{britton2006generating}
T.~Britton, M.~Deijfen, and A.~Martin-L{\"o}f.
\newblock Generating simple random graphs with prescribed degree distribution.
\newblock {\em Journal of Statistical Physics}, 124(6):1377--1397, 2006.

\bibitem{DiekmannEtal2013}
O.~Diekmann, H.~Heesterbeek, and T.~Britton.
\newblock {\em Mathematical Tools for Understanding Infectious Disease
  Dynamics}.
\newblock Princeton University Press, 2012.

\bibitem{Durrett2007}
R.~Durrett.
\newblock {\em {Random graph dynamics}}.
\newblock Cambridge University Press, 2006.

\bibitem{ErdosEtal1959}
P.~Erd{\H o}s and A.~R{\'e}nyi.
\newblock On random graphs, i.
\newblock {\em Publicationes Mathematicae (Debrecen)}, 6:290--297, 1959.

\bibitem{Grassberger2013}
P.~Grassberger.
\newblock Two-dimensional sir epidemics with long range infection.
\newblock {\em Journal of Statistical Physics}, 153(2):289--311, 2013.

\bibitem{Grimmett99}
G.~Grimmett.
\newblock {\em Percolation}, volume 321.
\newblock Springer-Verlag, Berlin, second edition, 1999.

\bibitem{Guttorp1991}
P.~Guttorp.
\newblock {\em Statistical inference for branching processes}, volume 122.
\newblock Wiley-Interscience, 1991.

\bibitem{Hofs14a}
R.~{\SortNoop{Hofstad}}van~der Hofstad.
\newblock {\em Random graphs and complex networks}, volume~1.
\newblock Cambridge University Press, 2016.

\bibitem{Holm2002sverige}
E.~Holm.
\newblock {\em The SVERIGE spatial microsimulation model: content, validation,
  and example applications}.
\newblock Department of Social and Economic Geography, Univ., 2002.

\bibitem{Jagers75}
P.~Jagers.
\newblock {\em Branching Processes with Biological Applications}.
\newblock Wiley, New York, 1975.

\bibitem{StanfordData}
J.~Leskovec and A.~Krevl.
\newblock Snap datasets: Stanford large network dataset collection.
\newblock {\em URL http://snap. stanford. edu/data}, 2014.

\bibitem{MalmrosEtal2016}
J.~Malmros, F.~Liljeros, and T.~Britton.
\newblock Respondent-driven sampling and an unusual epidemic.
\newblock {\em Journal of Applied Probability}, 53(02):518--530, 2016.

\bibitem{MatlabCFT}
The MathWorks, Inc., Natick, Massachusetts, United States.
\newblock {\em MATLAB and Curve Fitting Toolbox Release 2017a}.

\bibitem{Mollison1977}
D.~Mollison.
\newblock {Spatial contact models for ecological and epidemic spread}.
\newblock {\em J. R. Stat. Soc. Ser. B Stat. Methodol.}, 39(3):283--326, 1977.

\bibitem{MolloyEtal1995}
M.~Molloy and B.~Reed.
\newblock A critical point for random graphs with a given degree sequence.
\newblock {\em Random structures \& algorithms}, 6(2-3):161--180, 1995.

\bibitem{Newm02}
M.~E.~J. Newman.
\newblock Spread of epidemic disease on networks.
\newblock {\em Phys. Rev. E}, 66(1):016128, 11, 2002.

\bibitem{Reppell2014}
M.~Reppell, M.~Boehnke, and S.~Z{\"o}llner.
\newblock The impact of accelerating faster than exponential population growth
  on genetic variation.
\newblock {\em Genetics}, 196(3):819--828, 2014.

\bibitem{Tamhane2000}
A.~C. Tamhane and D.~D. Dunlop.
\newblock {\em Statistics and Data Analysis}.
\newblock Prentice-Hall, 2000.

\bibitem{Tolle2003}
J.~Tolle.
\newblock Can growth be faster than exponential, and just how slow is the
  logarithm?
\newblock {\em The Mathematical Gazette}, 87(510):522–525, 2003.

\bibitem{Trapman2010}
P.~Trapman.
\newblock The growth of the infinite long-range percolation cluster.
\newblock {\em Ann. Probab.}, 38(4):1583--1608, 2010.

\bibitem{Trapman2016}
P.~Trapman, F.~Ball, J.-S. Dhersin, V.~C. Tran, J.~Wallinga, and T.~Britton.
\newblock Inferring r0 in emerging epidemics—the effect of common population
  structure is small.
\newblock {\em Journal of The Royal Society Interface}, 13(121):20160288, 2016.

\bibitem{ViboudEtal2016}
C.~Viboud, L.~Simonsen, and G.~Chowell.
\newblock A generalized-growth model to characterize the early ascending phase
  of infectious disease outbreaks.
\newblock {\em Epidemics}, 15:27--37, 2016.

\bibitem{Wallace1991}
R.~Wallace.
\newblock Traveling waves of hiv infection on a low dimensional
  ‘socio-geographic’network.
\newblock {\em Social science \& medicine}, 32(7):847--852, 1991.

\bibitem{watts1998collective}
D.~J. Watts and S.~H. Strogatz.
\newblock Collective dynamics of'small-world'networks.
\newblock {\em nature}, 393(6684):440, 1998.

\end{thebibliography}

\appendix

\noindent\makebox[\linewidth]{\rule{\textwidth}{0.4pt}}
\begin{raggedright}
\section{Derivation of the Expected Reproduction Number for the Restricted Epidemic on the Configuration Model}
\end{raggedright}
\label{sec:derivationrepno}

We now study the expected reproduction number for a restricted epidemic on the configuration model. We assume that the branching process is valid and thus that each vertex produces offspring independently,  but drawn from the same distribution $L$. We focus on studying the expectation of $L$ and derive it based on the degree distribution $\Zt$ (see \refSubsec{epidemics}).

For maximum bound we consider $\Zt=k$ given and then $k-1$ edges can carry the infection on (taking into account that the infection cannot spread back on the edge that it arrived on). For each such stub consider the event $A_{i}=$ ``edge $i$ carries the infection on'' (with $i=1, 2, \dots, k-1$) and define the indicator variable\vspace{-2mm}
\begin{equation*}
	\One_{\{A_{i}\}}=
	\begin{cases}
		1 & \text{if } A_{i},\\
		0 & \text{otherwise}.
	\end{cases}\vspace{-2mm}
\end{equation*}
Then the number of offspring $L$ conditioned on $\Zt{=}k$ becomes\vspace{-2mm}
\begin{equation*}
	L\given \Zt{=}k\;\;=\;\;\sum\limits_{i=1}^{k-1}\One_{\{A_{i}\}}
\end{equation*}

There are $c$ attempts at carrying the infection on and for each one a neighbour is selected uniformly at random, with replacement. Thus $\Pr(A_{i}) = 1-\Pr({A_{i}}^{\complement})=1-\left(\frac{k-1}{k}\right)^{c}$, since for each of $c$ attempts the probability is $\frac{k-1}{k}$ that vertex $i$ is not selected to carry the infection on. Here ${A_{i}}^{\complement}$ indicates the complement of $A_{i}$. Thus\vspace{-2mm}
\begin{eqnarray*}
	\E[L\given\Zt=k] &=& \E\left[\sum\limits_{i=1}^{k-1}\One_{\{A_{i}\}}\right]
	= \sum\limits_{i=1}^{k-1}\E\left[\One_{\{A_{i}\}}\right]
	= \sum\limits_{i=1}^{k-1}\Pr(A_{i})\\
	&=& \sum\limits_{i=1}^{k-1}\left(1-\left(\frac{k-1}{k}\right)^{c}\right)
	= (k-1)\left(1-\left(1-\frac{1}{k}\right)^{c}\right)\vspace{-2mm}
\end{eqnarray*}
and so\vspace{-2mm}
\begin{eqnarray*}
	\E\left[L\given\Zt\right] &=& \left(\Zt-1\right)\left(1-\left(1-\frac{1}{\Zt}\right)^{c}\right),\vspace{-2mm}
\end{eqnarray*}
finally\vspace{-2mm}
\begin{equation}
	\E[L] = \E\left[\E\left[L\given\Zt\right]\right]=\E\left[\left(\Zt-1\right)\left(1-\left(1-\frac{1}{\Zt}\right)^{c}\right)\right].
\end{equation}

\end{document}